\documentclass[12pt]{article}
\usepackage{ifpdf}
\ifpdf
\usepackage[pdftex,bookmarks]{hyperref}%[pdftex,bookmarks]
\usepackage[pdftex]{graphicx}
\else
\usepackage{hyperref}
\usepackage[dvips]{graphicx}
\fi
\hypersetup{colorlinks=false,
	bookmarksopen,bookmarksnumbered,citecolor=blue,pdfstartview=FitH,
	pdfauthor = {Sergei M. Kuzenko and Simon J. Tyler},
	pdftitle = {Supersymmetric Euler-Heisenberg effective action: 
		Two-loop results},
	pdfsubject = {Physics},
	pdfkeywords = {SQED, supersymmetry},
	pdfcreator = {LaTeX with hyperref package},
	pdfproducer = {pdflatex}
}
\usepackage{latexsym}
\usepackage{amssymb,amsfonts,amsmath}
\usepackage{indentfirst}
\usepackage{bbm}
\usepackage[mathscr]{euscript}
\parskip=\medskipamount

\newcommand{\rmd}{{\rm d}}
\newcommand{\rme}{{\rm e}}
\newcommand{\rmi}{{\rm i}}
\renewcommand{\Re}{\ensuremath{\text{Re}}}
\renewcommand{\Im}{\ensuremath{\text{Im}}}
\def \intz{\int\!\!{\rmd}^8z}
\def \intc{\int\!\!{\rmd}^6z}
\newcommand{\eps}{\varepsilon}
\newcommand{\rmv}{{\rm v}}
\newcommand{\Bb}{{\bar B}}
\newcommand{\Db}{{\bar D}}
\newcommand{\Nb}{{\bar N}}
\newcommand{\Wb}{{\bar W}}
\newcommand{\cDb}{{\bar \cD}}
\newcommand{\cNb}{{\bar \cN}}
\newcommand{\cWb}{{\bar \cW}}
\newcommand{\half}{\ensuremath{\frac{1}{2}}}
\newcommand{\third}{\ensuremath{\frac{1}{3}}}
\newcommand{\quart}{\ensuremath{\frac{1}{4}}}

\newcommand{\pd}{\partial}
\newcommand{\Tr}{{\rm Tr}}
\newcommand{\tr}{{\rm tr}}

\newcommand{\ord}{\ensuremath{{\rm O}}}
\newcommand{\wrt}{with respect to~}
\def \cc {{\rm c.c.}}

\newcommand{\rhot}{\tilde{\rho}}
\newcommand{\scF}{\mathscr{F}}

\def \unren{{\rm unren}} 
\newcommand {\cA}{{\cal A}}

\newcommand {\cD}{{\cal D}}

\newcommand {\cF}{{\cal F}}

\newcommand {\cN}{{\cal N}}

\newcommand {\cQ}{{\cal Q}}

\newcommand {\cW}{{\cal W}}

\def\a{\alpha}
\def \bi{\bibitem}

\def\b{\beta}

\def\d{\delta}
\def\e{\epsilon}

\def\g{\gamma}
\def\G{\Gamma}

\def\j{\psi}
\def\k{\kappa}
\def\l{\lambda}

\def\p{\pi}
\def\q{\theta}
\def\r{\rho}
\def\s{\sigma}

\def\x{\xi}
\def\z{\zeta}
\def\D{\Delta}
\def\F{\Phi}

\def\O{\Omega}

\def\X{\Xi}
\def\tr{{\rm tr}}

\newcommand{\ad}{{\dot{\alpha}}}                           %new
\newcommand{\bd}{{\dot{\beta}}}                            %new
\newcommand{\gd}{{\dot{\gamma}}}                           %new
\newcommand{\sect}[1]{\setcounter{equation}{0}\section{#1}}

\newcommand{\be}{\begin{equation}}
\newcommand{\ee}{\end{equation}}
\newcommand{\bea}{\begin{eqnarray}}
\newcommand{\eea}{\end{eqnarray}}
\newcommand{\non}{\nonumber}

\def\ds1{\ensuremath{\mathbbm{1}}}
\def\double #1{#1{\hbox{\kern-2pt $#1$}}}
\begin{document}
%%%%%%%%%%%%%%%%
%%%%%%%%%%%%%%%%
%
%
\begin{titlepage}
\begin{flushright}
hep-th/0703269\\
March, 2007\\
\end{flushright}
\vspace{5mm}

\begin{center}
{\Large \bf  Supersymmetric Euler-Heisenberg effective action:\\
Two-loop results
}
\end{center}

\begin{center}
{\large  
Sergei M. Kuzenko\footnote{kuzenko@cyllene.uwa.edu.au}
and 
Simon J. Tyler\footnote{styler@physics.uwa.edu.au}
} \\
\vspace{5mm}

\footnotesize{
{\it School of Physics M013, The University of Western Australia\\
35 Stirling Highway, Crawley W.A. 6009, Australia}}  
~\\

\vspace{2mm}

\end{center}
\vspace{5mm}

\begin{abstract}
\baselineskip=14pt
%\noindent
The two-loop Euler-Heisenberg-type effective action for $\cN=1$ 
supersymmetric QED is computed within the background field approach. 
The background vector multiplet is chosen to obey the constraints
$D_\a W_\b = D_{(\a} W_{\b)} ={\rm const}$, but is otherwise completely arbitrary.
Technically, this calculation proves to be much more laborious  as compared 
with that carried out in hep-th/0308136 for $\cN=2$ supersymmetric QED,
due to a lesser amount of supersymmetry.
Similarly to  Ritus' analysis for spinor and scalar QED, 
the two-loop renormalisation is carried out using  proper-time
cut-off regularisation. 
A closed-form expression is obtained for 
the holomorphic sector of the two-loop effective action, which is 
 singled out by imposing  a relaxed 
super self-duality condition.
\end{abstract}
\vspace{1cm}

\vfill
\end{titlepage}

\newpage
\renewcommand{\thefootnote}{\arabic{footnote}}
\setcounter{footnote}{0}

%%%%%%%%%%%%%%%%%%%%%%%%%%%%%%%%%%%%%%%%%%%%%%%%%%%%%%%%%%%%%%%%%%%%%%%%%%%%%%%%
%%%%%%%%%%%%%%%%%%%%%%%%%%%%%%%%%%%%%%%%%%%%%%%%%%%%%%%%%%%%%%%%%%%%%%%%%%%%%%%%
\sect{Introduction}
In the mid-1930s,  two nonlinear generalisations of 
Maxwell's theory were introduced, 
the Born-Infeld action \cite{BI} and the Euler-Heisenberg
effective Lagrangian \cite{EH} (and its extension for scalar QED  \cite{W}).
Although these models were soon abandoned by their creators, 
their impact on the subsequent development of theoretical high-energy physics
has been profound. In particular,
the Born-Infeld action emerged naturally in string theory \cite{FT}
(see \cite{Tseytlin} for a review) 
as the bosonic sector of the vector Goldstone multiplet action for partial 
supersymmetry breaking \cite{BG}, and  as an example of 
self-dual models for nonlinear electrodynamics \cite{GR,GZ} 
(see \cite{KT} for a review and supersymmetric extensions).
As for the effective theories put forward in \cite{EH,W},
after Schwinger applied his functional techniques \cite{Schwinger} to 
re-derive and extend the results of  \cite{EH,W}, the Euler-Heisenberg 
Lagrangian has become a paradigm for practically all developments 
related to the evaluation and analysis of low-energy effective actions in 
quantum field theory, quantum gravity and string theory 
(for a review of Euler-Heisenberg effective Lagrangians, see \cite{Dunne}). 

The one-loop results for spinor and scalar QED \cite{EH,W,Schwinger} 
were extended in 1975 by Ritus to the two-loop approximation
\cite{Ritus}. Further analysis at two loops 
was carried out by many groups using various techniques,
see, e.g., \cite{DR,RSS,FRSS,KS}. 
In the supersymmetric case, the two-loop Euler-Heisenberg-type effective action 
has only been computed  for $\cN=2$ supersymmetric QED (SQED) 
\cite{KM}, using the covariant supergraph techniques formulated in \cite{KM2}.
The present paper is aimed at extending the results of \cite{KM} 
to the case of $\cN=1$ SQED.

By `supersymmetric Euler-Heisenberg action' we mean a sector of the 
low energy effective action of the form 
\be
\G = \frac{1}{e^2}  \int{\rm d}^6 z \,W^2
+ \int{\rm d}^8 z \, W^2{\bar W}^2 
\, \O\Big( D^2 W^2,  {\bar D}^2  {\bar W}^2 \Big)~.
\label{structure}
\ee
Such a  functional form is characteristic of the supersymmetric Born-Infeld
action \cite{CF,BG} and, more generally,   self-dual models for nonlinear 
supersymmetric electrodynamics \cite{KT}.
To compute the above sector of the effective action within the background field method, 
it is sufficient to make use of a  constant background vector multiplet constrained by 
\bea
D_\a W_\b = D_{(\a} W_{\b)} ={\rm const}~.
\eea
Since such a vector multiplet is a solution to the equations of motion for
any action functional $\G[W_\a, {\bar W}_\ad ]$, the action (\ref{structure}) 
is independent of the choice of gauge fixing in path integral.

This paper is organised as follows.
In section 2 we provide the necessary background field setup for 
$\cN=1$ SQED and, for a special background vector multiplet, 
express the matter propagators in terms of a single 
background-dependent Green's function for which an exact expression is known.
The one-loop effective action for $\cN=1$ SQED is reviewed in section 3.
Section 4 is the centre of this paper, and is devoted to the evaluation 
of the two-loop quantum corrections. 
Renormalisation of the previous sections' results is discussed in section 5.
In section 6 we derive a closed-form expression for 
a special holomorphic sector of the two-loop effective action.
${}$For this paper to be self-contained, we also included two technical 
appendices. Appendix A contains the expressions for the exact propagators 
in the presence of a constant background vector multiplet. 
Appendix B contains a simple derivation of the one-loop K\"ahler potential 
and chiral two-point function in the Fermi-Feynman gauge.

%%%%%%%%%%%%%%%%%%%%%%%%%%%%%%%%%%%%%%%%%%%%%%%%%%%%%%%%%%%%%%%%%%%%%%%%%%%%%%%%
\sect{Background field setup}
The classical action for $\cN=1$ SQED is 
\bea
S_{\rm SQED} = 
 \frac{1}{e^2} \int {\rm d}^6 z \, W^\a W_\a 
&+&\int {\rm d}^8 z \, \Big( \bar{Q}_+ {\rm e}^V Q_+ 
+ \bar{Q}_- {\rm e}^{-V} Q_- \Big)
\non \\
&+& \Big(
m  \int {\rm d}^6 z \, Q_+ Q_- + {\rm c.c.} \Big)~,
\label{n=1sqed-action0}
\eea
where the gauge field is described by a real unconstrained prepotential $V$ 
with\linebreak 
$W_\a = -(1/8){\bar D}^2D_\a V$ its gauge invariant field strength.
The supersymmetric matter is realised in terms of 
chiral superfields $Q_+$ and $Q_-$ of charge $+1$ and $-1$ respectively. 

It is instructive to compare the action (\ref{n=1sqed-action0})
with that
 for $\cN=2$ SQED:
\bea
S_{\rm SQED}^{\cN=2} &=& \frac{1}{e^2} \int {\rm d}^8 z \, {\bar \F} \F
+  \frac{1}{e^2} \int {\rm d}^6 z \, W^\a W_\a \non \\
&& +\int {\rm d}^8 z \, \Big( \bar{Q}_+ {\rm e}^V Q_+ 
+ \bar{Q }_- {\rm e}^{-V} Q_- \Big)
+ \Big(
 \int {\rm d}^6 z \, \F \,Q_+  Q_- + {\rm c.c.} \Big)~,
\label{n=2sqed-action0}
\eea
with $\F$ a neutral chiral superfield. 
Here the dynamical variables $\F$ and $V$ realise an
abelian $\cN=2$ vector multiplet, while 
the superfields $Q_+$ and  $Q_-$ constitute  
a massless Fayet-Sohnius hypermultiplet.
The case of a massive hypermultiplet is obtained 
from (\ref {n=2sqed-action0}) by the shift
$\F \to \F+ m$. 

One can see that 
the classical action of $\cN=1$ SQED, eq.   (\ref{n=1sqed-action0}),
is obtained from (\ref {n=2sqed-action0})
by discarding $\F$ as a dynamical variable, and instead
`freezing' $\F$ to a constant value $m$.
This also holds in quantum theory at the one-loop level. 
Specifically, if $\G_{\cN=1}^{(1)}
[W]$ and 
$\G_{\cN=2}^{(1)}
[W,\F]$ are the vector multiplet 
sectors of the one-loop effective actions for $\cN=1$
and $\cN=2$ SQED, respectively, then they are related to each other as follows:
$\G_{\cN=1}^{(1)} [W] = \G_{\cN=2}^{(1)} [W,m]$.
However, this simple correspondence breaks down already at two loops, 
due to the presence of additional supergraphs 
(involving internal $\F\bar \F$ lines) 
in the $\cN=2$ case.  This has the dramatic implication  
that the two-loop Euler-Heisenberg-type action for $\cN=1$ SQED 
is much more difficult to evaluate than the $\cN=2$  case 
\cite{KM}.

To quantise the theory (\ref{n=1sqed-action0})
within the background field formulation 
we first rewrite the action in terms of gauge covariantly chiral superfields
$\cQ_\pm$ and their conjugates. 
\bea
S_{\rm SQED} = \frac{1}{e^2} \int {\rm d}^6 z \, W^\a W_\a 
&+&\int {\rm d}^8 z \, \Big( \bar{\cQ}_+ \cQ_+ 
+ \bar{\cQ}_-  \cQ_- \Big) \non \\
&+& \Big(m  \int {\rm d}^6 z \, \cQ_+ \cQ_- + {\rm c.c.} \Big)~,
\label{n=1sqed-action1}
\eea         
where $\cQ_\pm$ satisfy the constraints
${\bar \cD}_\ad \cQ_\pm =0$,
with the  gauge covariant derivatives $\cD_A = (\cD_a, \cD_\a , {\bar \cD}^\ad ) 
= D_A +{\rm i}\cA_A(z)$ obeying the algebra
\bea\label{N=1cov-der-al}
	\left\{\cD_\a,\cD_\b\right\} &=& 
	\left\{\bar\cD_{\ad},\bar\cD_{\bd}\right\}=0~, \hspace{35pt}
	\left\{\cD_\a,\bar\cD_{\bd}\right\} = - 2\rmi \, \cD_{\a\bd}~, \non \\\non
	\left[{\cD_\a},{\cD_{\b\bd}}\right] &=& 2\rmi\eps_{\a\b}\,\cWb_{\bd}~, 
	\hspace{59pt}	 
	\left[{\bar\cD_{\ad}},{\cD_{\b\bd}}\right] = 2\rmi\eps_{\ad\bd}\,\cW_\b~,\\
	\left[{\cD_{\a\ad}},{\cD_{\b\bd}}\right] &=& \rmi\cF_{\a\ad, \b\bd} 
	= - \eps_{\a \b}\, {\bar \cD}_{\ad} {\bar \cW}_{\bd}
	-\eps_{\ad\bd} \,\cD_\a \cW_\b~.
\eea
Here the action of $\cW_\a $ on $\cQ_\pm$ is defined as follows:
$\cW_\a \cQ_\pm = \pm W_\a \cQ_\pm$.

In accordance with  the $\cN=1$ 
background field formulation \cite{GRS,GGRS}, we 
split the dynamical variables into  background 
and quantum,
\bea
 \cQ_\pm  ~ \to ~ \cQ_\pm  +q_\pm ~, 
\qquad   \quad
\cD_\a ~ \to ~ {\rm e}^{-\hat{e}\,v} \, \cD_\a \, {\rm e}^{\hat{e}\, v}~, 
\qquad
{\bar \cD}_\ad ~ \to ~ {\bar \cD}_\ad~,
\label{bq-splitting}
\eea
with lower-case letters used for 
the quantum superfields. 
Here $\hat e$ is the charge operator, $\hat{e} q_\pm = \pm e q_\pm$.
The quantum matter superfields $q_\pm$ are 
background covariantly chiral, 
\be
{\bar \cD}_\ad q_\pm =0~.
\ee
In this paper, we are mainly interested in the slowly varying part of the 
effective action that solely depends on the vector multiplet. For this it is 
sufficient to only consider a background that satisfies
\be 
\pd_a W_\b = D^\b W_\b=0~, \qquad \cQ_\pm =0~. \ee
Upon quantisation in the Fermi-Feynman gauge, 
we end up with the following action to be used for 
loop calculations 
\bea
S_{\rm quantum} = -\frac{1}{2} \int {\rm d}^8 z \, v \Box v 
&+&\int {\rm d}^8 z \, \Big( \bar{q}_+ {\rm e}^{ev} q_+ 
+ \bar{q}_- {\rm e}^{-e v} q_- \Big)
\non \\ &+& 
 \Big(m  \int {\rm d}^6 z \, q_+ q_- + {\rm c.c.} \Big)~.
\label{n=1sqed-action2}
\eea
${}$From here we can read off the propagators in the standard manner
\bea\label{propagators} \rmi\left<v(z)v(z')\right> &=& -G_0(z,z') ~,\non\\
	 \rmi\left<q_+(z)q_-(z')\right> &=& -m G_+(z,z') 
	 = \frac{m}{4}\bar\cD^2 G(z,z') ~,\non\\
	 \rmi\left<q_+(z)\bar q_+(z')\right> &=& G_{+-}(z,z') 
	 = \frac{1}{16}\bar\cD^2\cD^2 G(z,z')~, \non\\
	 \rmi\left<\bar{q}_-(z)q_-(z')\right> &=& G_{-+}(z,z') = G_{+-}(z',z) ~.
\eea
The above matter propagators are expressed 
via the Green's function $G(z,z')$ which
satisfy the equation
\be
(\Box_\rmv-m^2)G(z,z')=-
\delta^8(z-z')~, \qquad
\Box_\rmv = \cD^a\cD_a-\cW^\a\cD_\a+\cWb_\ad\cDb^\ad~,
\ee
and is characterised by the proper-time 
representation (\ref{greensfn}) and (\ref{heatk}).
The proper-time representation for the free,  massless Greens function, 
$G_0(z,z')$, which determines the gauge field propagator is
\bea 
G_0(z,z') &=& \rmi\int\limits_0^\infty\!\rmd s\, K_0(z,z'|s)\,
{\rm e}^{-\e s}~, \qquad \e \to +0 ~, \non \\
 K_0(z,z'|s)&=&K_{\rm bos}(\rho|s)\delta^4(\z)
=-\frac{\rmi}{(4\pi s)^2}\rme^{\rmi\rho^2/4s} \delta^4(\z)~,
\label{G0} 
\eea
with $K_{\rm bos}$  the bosonic heat kernel, and the two point functions 
$(\r^a,\z^\a,{\bar\z}_\ad)$ defined in (\ref{zeta}).
The interactions are easily read from (\ref{n=1sqed-action2}) 
by expanding in the quantum fields.

%%%%%%%%%%%%%%%%%%%%%%%%%%%%%%%%%%%%%%%%%%%%%%%%%%%%%%%%%%%%%%%%%%%%%%%%%%%%%%%%
\sect{One-loop effective action}
\label{sect1loop}
Although the one-loop Euler-Heisenberg effective action for SQED has been 
calculated in many other places \cite{SY,Ohrn,MG,PB,BKT,KM}, 
we will repeat it here for the sake of completeness and in order to establish 
some notation.  Its formal representation is 
(see \cite{BK} for an introduction to heat kernel techniques in superspace)
\be
\Gamma^{(1)}_\unren=-\rmi \, \Tr_+\ln G_+ = 
	-\rmi\int\limits_{s_0}^\infty\frac{\rmd s}s\,
	\Tr_+ K_+(s)\rme^{-\rmi (m^2 -{\rm i}\e )s}~,
\ee
where we have introduced a proper-time cut-off to regularise 
UV divergences.  We note that 
the standard $\rmi\eps$ prescription (see also eq.  (\ref{greensfn}))
 is equivalent to having $\Im(s_0)<0$.

The above functional trace of the chiral heat kernel (\ref{chheatk}) 
is defined by
\be \Tr_+K_+(s)=\intc \,K_+(z,z|s) ~,\ee
so 
the evaluation of $\Gamma^{(1)}$ reduces  
to finding the coincidence limit of $K_+(z,z'|s)$.
In accordance with the results listed in Appendix \ref{app-superprop}, 
eq. (\ref{UofsAction}), it follows that
\be 
\z(s)^2\Big|_{\z\to0} =2 W^2 \,\frac{\cos(sB)-1}{B^2}~.
 \ee
This factor of $W^2$ then prevents any further contributions coming from 
the action of $U(s)$, thus in the coincidence limit the exponential 
and the parallel propagator go to unity.
In the above we have introduced the notation
\be B^2=\half\tr{N}^2,\quad N_\a^{~\b}=D_\a W^\b~;\qquad
	\Bb^2=\half\tr{\Nb}^2,\quad \Nb^\ad_{~\bd}=\Db^\ad \Wb_\bd~,
\ee
and for the on-shell backgrounds that we are using, the above definitions imply
\be\label{traceN} \tr N^{2n}=2B^{2n}~,\qquad \tr N^{2n+1}=0~,
\ee
formulae that will be repeatedly used in the following section.
These objects also appear in the eigenvalues of $F=F_a^{~b}$ 
which are equal to
$\pm\l_+$ and $\pm\l_-$, where
\be
\label{Feigen} \l_\pm=\frac\rmi2(B\pm\Bb)~. 
\ee
This then allows the calculation of the determinant
\be
\label{lor.det} 
\sqrt{\det\left(\frac{2sF}{{\rm e}^{2sF}-1}\right)}
=\frac{s\l_+}{\sinh(s\l_+)}\frac{s\l_-}{\sinh(s\l_-)} 
	=-\frac{s^2}2\frac{B^2-\Bb^2}{\cos(sB)-\cos(s\Bb)}~.
	\ee
So, after a small amount of algebra to separate off the term that leads to the 
UV divergence, the heat kernel reduces to
\be 
K_+(s)=\frac{\rmi }{(4\p)^2} \,W^2\left(1+\frac{\Bb^2(1-\cos(sB))
-B^2(1-\cos(s\Bb)}{B^2(\cos(sB)-\cos(s\Bb))}\right)~. 
\label{chk}
\ee
Then, following \cite{BKT}, we note that the quartic and higher order terms
on the right of (\ref{chk}) contain a factor of $\Bb^2$,
and the latter can be represented as
$\Bb^2=\quart \Db^2\Wb^2$ for the background chosen.
This allows us to bring 
the unrenormalised one-loop effective action
to the form:
\be
\label{1loopunren} 
\begin{split}
	\Gamma_\unren^{(1)}=& \frac{1}{(4\pi)^2}\int\limits_{s_0}^\infty\!
	\frac{\rmd s}{s}\,
	\rme^{-\rmi (m^2 -{\rm i}\e)s} \intc W^2 
	\\
	+&\frac{1}{(4\pi)^2}\intz W^2\bar{W}^2\int\limits_0^\infty\!\rmd s
	\frac{\Bb^2\left(\cos(sB)-1\right)-B^2\left(\cos(s\Bb)-1\right)}
	{s\;B^2\Bb^2	\left(\cos(sB)-\cos(s\Bb)\right)}\,
	\rme ^{-\rmi (m^2-{\rm i}\e)s}~.
	\end{split}
\ee
The first term is obviously UV divergent as $s_0\to0$ and 
is absorbed into the renormalisation of $e^2$.
This is discussed in detail in section \ref{renorm}.

\begin{figure}[t]
\begin{minipage}[t]{.60\textwidth}\centering
\includegraphics{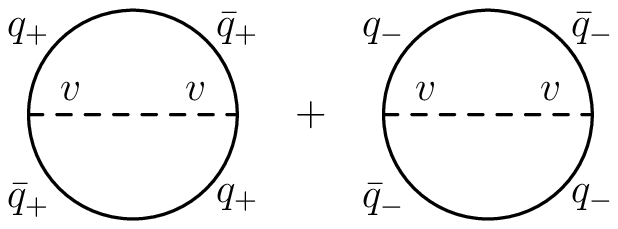}
\caption{Two-loop supergraph I}\label{2loopgraphI}
\end{minipage}%
\begin{minipage}[t]{.4\textwidth}\centering
\includegraphics{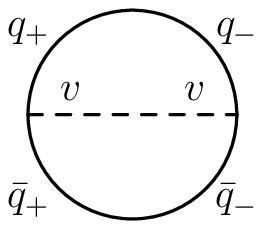}
\caption{Two-loop supergraph II}\label{2loopgraphII}
\end{minipage}%
\end{figure}

%%%%%%%%%%%%%%%%%%%%%%%%%%%%%%%%%%%%%%%%%%%%%%%%%%%%%%%%%%%%%%%%%%%%%%%%%%%%%%%%
\sect{Two-loop quantum corrections}
\label{sect2loop}
We now come to the central calculation of this paper, the two-loop quantum 
correction to the effective action.
There are two non-zero\nolinebreak
\footnote{There is a third 1PI supergraph, the so-called `figure eight' graph, 
whose contribution is easily seen to be zero in the Fermi-Feynman gauge.}, 
two-loop 1PI supergraphs, 
as shown in figures \ref{2loopgraphI} and \ref{2loopgraphII}.
The first diagram contributes
\be
\label{2I} 
\Gamma^{(2)}_{\rm I}=\frac{e^2}{2^8}\intz\!\!\intz'\,
	 G_0(z,z')\cD^2\bar\cD^2G(z,z')\,\bar\cD^2\cD^2G(z',z)~,  
\ee
whilst the contribution from the second diagram is
\be
\label{2II} 	
\Gamma^{(2)}_{\rm II}=-\frac{e^2}{2^4} m^2\intz\!\!\intz'~
	  G_0(z,z')\bar\cD^2G(z,z')\,\cD^2G(z',z)~.
\ee
Inserting the proper-time representations 
for the Green's functions into (\ref{2I}) gives
\be 
\G^{(2)}_{\rm I}=-\rmi e^2\intz\rmd^8z'\!\!\int\limits_0^{\infty}\! 
	\rmd s\rmd t\rmd u\,{K_0}(z,z'|u)K_{-+}(z,z'|s)
	K_{-+}(z',z|t)\rme^{-\rmi (m^2-{\rm i}\e)(s+t)}~,
\label{qc}
\ee
and similarly for $\Gamma^{(2)}_{\rm II}$.
We should emphasise that here we only collect the unregulated 
two-loop quantum corrections. The issues of regularisation 
and renormalisation will be discussed in detail in the following section. 

Before plunging into actual calculations,
it is instructive to compare  the quantum correction
(\ref{2I}) with its counterpart in the case of $\cN=2$ SQED \cite{KM}.
As mentioned in Section 2, in the $\cN=2$ case there 
is a third diagram which can be combined with the 
first to give a dramatic simplification.  
Their combined total contribution
can be obtained from (\ref{qc})
by replacing 
\be
K_{-+}(z,z'|s)~\to~ K_{-+}(z,z'|s) -K_{+-}(z,z'|s)
= \frac{1}{16}[\cD^2,\cDb^2]K(z,z'|s)~,
\label{cq2}
\ee
where we have used the identity \cite{KM2}
\be
 \cDb'^2K(z,z'|s)=\cDb^2K(z,z'|s)~.
\ee
Then, the structure of the resulting quantum correction is such that
$K_{-+}(z',z|t)$ can be equivalently replaced 
\bea
2K_{-+}(z',z|t) ~\to~ K_{-+}(z',z|t)-K_{+-}(z',z|t)
=-\frac{1}{16}[\cD^2,\cDb^2]K(z',z|t)~.
\label{cq3}
\eea
In the Grassmann coincidence limit, it can be shown \cite{KM} that
\be
[\cD^2,\cDb^2]K(z,z'|s)\big|_{\z=0} \propto W_\a {\bar W}_\ad~,
\ee
and similarly for the expression in (\ref{cq3}). As a result, 
the $\cN=2$ counterpart of (\ref{qc}) contains a factor of $W^2 {\bar W}^2$
in the integrand. It is  this technical property that allows the dramatic 
simplification of all further calculations \cite{KM}.
This has no analogue in the case of $\cN=1$ SQED.

So we now continue with the evaluation of $\G^{(2)}_{\rm I}$ by 
integrating over the primed Grassmann coordinates with the help of the 
delta function contained in the vector heat kernel.
We also shift the remaining spatial integration variables 
via the rule $\{x,x'\}\to\{x,\rho\}$ to yield
\be 
\G^{(2)}_{\rm I}=\rmi e^2\intz\rmd^4\r\!\!\int\limits_0^{\infty}\!
	\rmd s\rmd t\rmd u\,{K}_{\rm bos}(\rho|u)K_{-+}(z,z'|s)
	K_{-+}(z',z|t)\rme^{-\rmi (m^2-{\rm i}\e)(s+t)}\Big|_{\z=0}~.
\ee
We see that to find $\G^{(2)}_{\rm I}$ we first have to calculate the 
antichiral-chiral heat kernel, defined below,
in the Grassmann coincidence limit $(\zeta_\a,~\bar\z_\ad)\to 0$.

As was demonstrated in \cite{KM}, for the constant, on-shell backgrounds 
that we are considering, the antichiral-chiral heat kernel
can be obtained by taking derivatives of the heat kernel (\ref{heatk}),
\bea
\label{ac.chk} 
K_{-+}(z,z'|s)&=&\frac{1}{16}\cD^2\cDb'^2K(z,z'|s) \non \\
	&=&\frac{-\rmi}{(4\pi s)^2}\sqrt{\det
	\left(\frac{2sF}{{\rm e}^{2sF}-1}\right)}~U\!(s)~
	\rme^{\frac\rmi{4}\rhot\,F\!\coth(sF)\,\rhot+R(z,z')}I(z,z')~,
\eea
where 
\be\non R(z,z')=\third(\zeta^2\bar\z\cWb-\bar\z^2\zeta\cW)
	-\frac{\rmi}2\rhot^a(\cW\sigma_a\bar\z+\zeta\sigma_a\cWb)
	-\frac{\rmi}{12}\rhot_{\a\ad}(\zeta^\a\bar\z\bar\cD\cWb^{\ad}+
	5\bar\z^{\ad}\zeta\cD\cW^\a)~, \label{R}
\ee
and $\rhot$ is a two-point variable that is a antichiral in $z$ and chiral 
in $z'$,
\be\label{rhot}
	\rhot^a=\rho^a-i\zeta\sigma^a\bar\z, \quad 
	\cD_\b\rhot^a=\bar\cD'_{\bd}\rhot=0~. 
\ee
If we use the notation $\Psi(s)\equiv U\!(s)\Psi U\!(-s)$ 
for proper-time dependent variables we see
that the action of $U\!(s)$ in (\ref{ac.chk}) can be summarised by modifying 
the exponential to be
\be\non
 \rme^{\frac\rmi4\rhot(s)\,F\!\coth(sF)\,\rhot(s)+R(z,z')
 +\int_0^s\!\rmd t \left(R'(t)+\X(t)\right)} ~,
\ee
where the action of $U\!(s)$ on $\rho$, $\z$, $W^\a$ and $I(z,z')$ 
is displayed in (\ref{UofsAction}) and $\X$ is defined in (\ref{Xi}).
The reason for writing $R(s)$ in the convoluted way above becomes clear when 
we note that $R(z,z')|_{\z\to0}=0$ and
\be R'(t)+\X(t)=\rmi U(t)\left(2\zeta^2\Wb^2-\zeta^2\bar\zeta\Nb\Wb
	+\frac\rmi{2}(\zeta\rho\Nb\Wb + \zeta N\rho\Wb) \right)U(-t)~. 
\ee
It is now a straightforward but tedious task to take the Grassmann 
coincidence limit. 
We found it simplest to perform this limit by looking at the first 
and last term in the exponential separately.
Using (\ref{traceN}) and writing 
\be\label{FasN}
F_{\a\ad \b\bd}=\s^a_{\a\ad}\s^b_{\b\bd} F_{ab}=\rmi(\eps_{\a\b}\Nb_{\ad\bd}
+\eps_{\ad\bd}\Nb_{\a\b})~,
\ee
to assist in taking the traces, we get
\bea 
&&K_{-+}(z,z'|s)\Big|_{\z=0}=-\frac{\rmi}{(4\pi s)^2}{
\det\!\left(\frac{2sF}{{\rm e}^{2sF}-1}\right)}^{\!\half} \non \\
&& \qquad  \times 	\exp \Big\{
{\frac\rmi{4}\rho\,F\!\coth(sF)\,\rho 
-\rmi W^\b \rho_a f^{a}_{~\b\bd}(s)\Wb^\bd-\rmi W^2\Wb^2 f(s)}
\Big\}	I(z,z')\Big|_{\z=0}~ ,\qquad\qquad\quad
\eea  
where
\bea 
	f^a_{~\b\bd}(s)&=&\half(\ds1-\coth(s F))^a_{~b}\Big(
	\frac{\rme^{-\rmi sN}-\ds1}{N}\s^b\rme^{-\rmi s\Nb}
	+\s^b\frac{\rme^{-\rmi s\Nb}-\ds1}{\Nb}\Big)_{\b\bd}~,\\
	f(s)&=&4\frac{B\sin(sB)\sin^2(s\Bb/2)-\Bb\sin(s\Bb)\sin^2(sB/2)}
	{B^2\Bb^2(\cos(sB)-\cos(s\Bb)} \non~.
\eea

The coincidence limit of $K_{-+}(z',z|t)$ is simply obtained 
from the above result via
the obvious replacements $z\leftrightarrow z'$ and $s\to t$.  Then by
pushing the parallel displacement operator through to the left we can combine 
the two heat kernels  to get
\bea 
\G^{(2)}_{\rm I}=&&\frac{e^2}{(4\pi)^6}\intz\rmd^4\r\!\!\int\limits_0^{\infty}\!
	\frac{\rmd s\rmd t\rmd u}{s^2t^2u^2}P_+P_- \\\non
	&&\times\rme^{\frac\rmi4\rho A\rho
	-\rmi W^\b \rho_a \left(f^{a}_{~\b\bd}(s)-f^{a}_{~\b\bd}(t)\right)\Wb^\bd
	-\rmi W^2\Wb^2 \left(f(s)+f(t)\right)}\rme^{-\rmi (m^2-{\rm i}\e)(s+t)}~,
\eea
where the parallel displacement operators have annihilated each other, 
in accordance with (\ref{par-prop}). 
Here we've introduced the notations
\bea P_\pm&=&\frac{s\l_\pm}{\sinh(s\l_\pm)}\frac{t\l_\pm}{\sinh(t\l_\pm)}~, \\
	A&=&F\coth(sF)+F\coth(tF)+\frac{1}{u} ~,
\eea
where the $P_\pm$ come from the determinant (\ref{lor.det}).
	
All $\rho$ dependence is now explicit in the exponential, 
so we can perform the gaussian 
integral to yield
\be 
	\frac1{(4\pi)^2}\int\!\!\rmd^4\r\, \rme^{\frac\rmi4\rho A\rho-
	\rmi W^\b \rho_a \left(f^{a}_{~\b\bd}(s)-f^{a}_{~\b\bd}(t)\right)\Wb^\bd}
	=\frac\rmi{\sqrt{\det A}}\rme^{-\rmi W^2\Wb^2\scF(s,t,u)}~,
\ee
where
\be \label{scFdef}
	\scF(s,t,u)=\quart\left(f^{a}_{~\g\gd}(s)-f^{a}_{~\g\gd}(t)\right) 
	(A^{-1})_a^{~b} \left(f_b^{~\gd\g}(s)-f_b^{~\gd\a}(t)\right)~,
\ee
and recalling the eigenvalues of $F$, (\ref{Feigen}), we obtain
\bea \frac1{\sqrt{\det A}}&=&\frac1{(a_++u^{-1})(a_-+u^{-1})}~,\\ 
a_\pm&=&\l_\pm\coth(s\l_\pm)+\l_\pm\coth(t\l_\pm)~. 
\eea
Equation (\ref{scFdef}) can be evaluated with the help of (\ref{traceN}), 
(\ref{FasN}) and the identity
\be 
\left(\coth(s\l_\pm)+1\right)\left(\coth(t\l_\pm)-1\right)=
-\frac{\rme^{\rmi\frac{B\pm\Bb}2(s-t)}}
{\sin(s\frac{B\pm\Bb}2)\sin(t\frac{B\pm\Bb}2)}~.
\ee
After some work it yields
\be
\scF(s,t,u)=\frac{\scF_+}{a_++u^{-1}}+\frac{\scF_-}{a_-+u^{-1}} ~,
\ee
with $\scF_+ \xrightarrow{\Bb \to -\Bb} \scF_-$ and, taking 
advantage of the integrands $s\leftrightarrow t$ symmetry,
\bea
\scF_+=&&2\frac{B^2\sin^2(\frac{s\Bb}2)+(B\leftrightarrow\Bb)
+2B\Bb\cos(s\frac{B+\Bb}2)\sin(\frac{sB}2)\sin(\frac{s\Bb}2)}%
	{B^2\Bb^2\sin^2(s\frac{B+\Bb}2)}\\\non
&&-2\frac{B^2\cos(B\frac{s-t}2)\sin(\frac{s\Bb}2)\sin(\frac{t\Bb}2)+
(B\leftrightarrow\Bb)+
	2B\Bb\cos(\frac{sB+t\Bb}2)\sin(\frac{tB}2)\sin(\frac{s\Bb}2)}
	{B^2\Bb^2\sin(s\frac{B+\Bb}2)\sin(t\frac{B+\Bb}2)}~.
\eea

Since $W_\a W_\b W_\g=0$ we get a simple, terminating expansion for the 
remaining exponential in $\G^{(2)}_{\rm I}$,
$$\rme^{-\rmi W^2\Wb^2(f(s)+f(t)+\scF(s,t,u))}
=1-\rmi \, W^2\Wb^2\Big(f(s)+f(t)+\scF(s,t,u)\Big)~.$$
Here the first term does not contribute to the Euler-Heisenberg sector of the 
effective action 
(it actually leads to  higher derivative quantum corrections),
so the first supergraph reduces to
\bea\label{2Ia}
	\G^{(2)}_{\rm I}=&&\frac{e^2}{(4\pi)^4}
	\intz W^2\Wb^2\!\!\int\limits_0^{\infty}\!
	\frac{\rmd s\rmd t\rmd u}{s^2t^2u^2}\frac{P_+}{a_++u^{-1}}
	\frac{P_-}{a_-+u^{-1}}\times\\\non
	&&\times\left(f(s)+f(t)+\frac{\scF_+}{a_++u^{-1}}
	+\frac{\scF_-}{a_-+u^{-1}}\right)
	\rme^{-\rmi (m^2 -{\rm i}\e) (s+t)}~.
\eea

The second supergraph is identical to one calculated in \cite{KM} 
so we just restate the result in our notation,
\be\label{2IIa}
\G^{(2)}_{\rm II}=\frac{e^2}{(4\pi)^4}\intz W^2\Wb^2\!\!\int\limits_0^{\infty}\!
\frac{\rmd s\rmd t\rmd u}{s^2t^2u^2}\frac{P_+}{a_++u^{-1}}\frac{P_-}{a_-+u^{-1}}
T(s,t)~	\rme^{-\rmi (m^2 - {\rm i}\e)(s+t)}~,
\ee
where 
\be T(s,t)=-\frac{8\rmi m^2}{B^2\Bb^2}\left(\sin^2(\frac{sB}2)
\sin^2(\frac{t\Bb}2)+s\leftrightarrow t\right)~. 
\ee

The two proper-time $u$-integrals in (\ref{2Ia}) and (\ref{2IIa}) 
can be performed in closed form and 
are identical to those considered by Ritus \cite{Ritus}.
Their direct evaluation gives 
\bea 
&&
\int\limits_{0}^{\infty}\frac{\rmd u}{u^2}\frac1{(a_++u^{-1})(a_-+u^{-1})}
		=\frac1{a_+-a_-}\ln\left(\frac{a_+}{a_-}\right)~, \\
&&
\int\limits_{0}^{\infty}\frac{\rmd u}{u^2}\frac1{(a_++u^{-1})(a_-+u^{-1})}
	\left(\frac{\scF_+}{a_++u^{-1}}+\frac{\scF_-}{a_-+u^{-1}}\right)\non \\
	&&\qquad =\frac1{a_+-a_-}\left(\frac{\scF_-}{a_-}-\frac{\scF_+}{a_+}\right)+
	 \frac{\scF_+-\scF_-}{(a_+-a_-)^2}\ln\left(\frac{a_+}{a_-}\right)~.
\eea

We can now write down the complete unrenormalised 2-loop effective action
\bea\label{unrenorm}
	\G_\unren^{(2)}=&&\frac{e^2}{(4\pi)^4}
	\intz W^2\Wb^2\!\!\int\limits_0^{\infty}\!
	\frac{\rmd s\rmd t}{s^2t^2}\frac{P_+P_-}{a_+-a_-}
	\left(\frac{\scF_-}{a_-}-\frac{\scF_+}{a_+}+\right.\\\non
	&&\left.+\left(f(s)+f(t)+T(s,t)+\frac{\scF_+-\scF_-}{a_+-a_-}\right)
	\ln\left(\frac{a_+}{a_-}\right)\right)
	\rme^{-\rmi (m^2 -{\rm i}\e) (s+t)}~.
\eea

%%%%%%%%%%%%%%%%%%%%%%%%%%%%%%%%%%%%%%%%%%%%%%%%%%%%%%%%%%%%%%%%%%%%%%%%%%%%%%%%
\sect{Renormalisation}\label{renorm}
As previously mentioned, we have regularised the divergences  
by using a proper-time cut-off.  
These cut-off dependent divergences are then removed in the standard way,
by adding counterterms to the original action.
Since the use of the background field method gives us the
freedom to rescale the quantum fields \cite{Abbott}, 
and gauge invariance implies that
the background gauge field $W^\a$ is not renormalised\nolinebreak
\footnote{Normally it is the combination $e V$ that is renormalisation invariant, 
but we have absorbed the charge into the field strength.},
the counterterm action takes the simple form
\be \label{CTact}
	 \frac{1}{e^2} (Z_e-1)
	\intc \,W^2 + (Z_Q-1) \left(m\intc \,q_+q_-+\cc\right)~.
\ee
We note that the first term above is proportional to the classical 
action, $\G^{(0)}=\frac{1}{e^2}\intc W^2$.
The counterterm coefficients are derived from the multiplicative 
renormalisation of charge and mass via
\bea  
e^2=Z_e e_0^2~, \qquad
m^2=Z_m m_0^2=Z_Q^{-2}m_0^2~, 
\eea
where we have used the fact that the $\cN=1$ nonrenormalisation theorem 
\cite{N=1non-ren,N=1non-ren2,GRS}
implies that $Z_m^\half \,Z_Q=1$.
The renormalisation constants are expanded with respect to the 
fine structure constant, $\a=e^2/8\p$,
\be 
Z_e= 1+ Z_e^{(1)}+Z_e^{(2)}+\ldots~,\qquad
Z_Q=1+ Z_Q^{(1)}+Z_Q^{(2)}+\ldots~. 
\ee
It is worth noting that in (S)QED an expansion in $\a$ is 
equivalent to the loop expansion. 

Each term in the loop expansion of the effective action is constructed from 
both the standard diagrams computed 
in the sections above and from diagrams with counterterm insertions.
There is a freedom in how much of the finite part of $\G_\unren^{(n)}$ is 
to be removed by the counterterm
contribution $\G_{\rm ct}^{(n)}$.
This corresponds to the freedom of choosing the finite part of the charge 
and matter renormalisation and
can be fixed by either choosing a consistent subtraction scheme, 
for example a (modified) minimal subtraction, or
by enforcing some renormalisation conditions.

We choose to work with physical parameters and thus calculate the counterterms 
using physical renormalisation conditions.
Following \cite{SVZ}, we define the physical charge squared as the inverse 
of the coefficient in front of the $W^2$ term.
This clearly leads to the correct charge in the gauge-matter coupling.  
The physical mass is harder to define from
within the Euler-Heisenberg sector of the effective action.  
The standard way to proceed is to use a separate
calculation of, for example, the K\"ahler potential and 
use the physical renormalisation conditions in that sector to 
find the correct mass renormalisation.  
This is done in appendix \ref{app.kahler}

First we examine the one-loop renormalisation.  
Adding the one-loop counterterm contribution to (\ref{1loopunren}) yields
\bea
\G^{(1)}&=& \G^{(1)}_{\rm ct} + \G_\unren^{(1)} 
=\frac1{e^2}\left(Z_e^{(1)}+\frac{\a}{2\p}E_1(\rmi m^2 s_0) \right)\intc \,W^2 
 \\ \non &&+ \frac{1}{(4\pi)^2}\intz \,W^2\bar{W}^2\int\limits_0^\infty\!\rmd s
\frac{\Bb^2\left(\cos(sB)-1\right)-B^2\left(\cos(s\Bb)-1\right)}{s\;B^2\Bb^2
\left(\cos(sB)-\cos(s\Bb)\right)}\,\rme ^{-\rmi (m^2 -{\rm i}\e)s}~,
\eea
where the exponential integral, $E_1$, 
is defined by \cite{AandS} 
\be E_n(z)=\int\limits_1^\infty\!\rmd t \frac{{\rm e}^{-z t}}{t^n}~,\quad
n=0,1,2,\ldots~,\quad \Re(z)>0~, \ee
with $E_1(z) = -\ln(z\rme^\g)+\ord(z)$ where $\g$ is the Euler-Mascheroni 
constant. It is clear that the renormalisation condition implies
\be 
Z_e^{(1)}=-\frac\a{2\p}E_1(\rmi m^2 s_0)~,
\ee 
so that the renormalised one-loop quantum correction is
\be\label{full1loop} 
\G^{(1)} = \frac{1}{(4\pi)^2}\intz \,W^2\bar{W}^2\int\limits_0^\infty\!\rmd s
\frac{\Bb^2\left(\cos(sB)-1\right)-B^2\left(\cos(s\Bb)-1\right)}{s\;B^2\Bb^2
\left(\cos(sB)-\cos(s\Bb)\right)}\,\rme ^{-\rmi (m^2 -{\rm i}\e)s}~.
\ee

Now we examine the two-loop renormalisation.
The two-loop counterterm contributions, read from (\ref{CTact}), are
\be \G^{(2)}_{\rm ct} = Z_e^{(2)}\, \G^{(0)}
+\rmi \,m^2 Z_Q^{(1)}\Big(\Tr_+G_+ + \Tr_-G_-\Big)~. \ee
This can be reduced to a more useful form by noting 
$\Tr_+G_+ = \Tr_-G_-$, see \cite{BK}, 
and that
\be
 \frac\pd{\pd m^2}\G^{(1)}_\unren 
 = -\rmi\frac\pd{\pd m^2}\int\limits_{0}^\infty\!\frac{\rmd s}s\Tr_+K_+ 
\rme^{-\rmi (m^2 -{\rm i}\e)s}
 =\rmi \, \Tr_+G_+ ~.
\ee
Then using $\G^{(1)}_\unren=\G^{(1)}-\G^{(1)}_{\rm ct}$ combined with the fact
\be \non
m^2\frac{\pd}{\pd m^2}E_1(\rmi m^2s_0)=-\rme^{-\rmi m^2 s_0}~,
\ee
we have
\be 
\G^{(2)}_{\rm ct} 
= \left(Z_e^{(2)}-\frac\a\p Z_Q^{(1)}\rme^{-\rmi m^2 s_0}\right)\G^{(0)}+
2Z_Q^{(1)}m^2\frac\pd{\pd m^2}\G^{(1)}~.
\ee 

A close examination of the proper-time integrand in the unrenormalised two-loop 
effective action (\ref{unrenorm}) 
shows that the only divergences that occur are in the 
$f(s)$ and $f(t)$ terms when $t$ or $s$ go to zero respectively.
Writing the unrenormalised result as
\be\label{unrenorm.1}
	\G^{(2)}_\unren=\frac{e^2}{(4\pi)^4}\intz \,W^2\Wb^2\!\!
	\int\limits_{s_0}^{\infty}\!\!
	\rmd s\rmd t ~ F(s,t)~\rme^{-\rmi (m^2 -{\rm i}\e)(s+t)}~,
\ee
we can separate off the divergent contribution 
by adding and subtracting the limit
\be\label{Ftilde}	
	\tilde F(s)\equiv\lim_{t\to 0}t F(s,t)
	=-2\frac{B^2-\Bb^2}{B^2\Bb^2}\frac{B\sin(sB)\sin^2(s\Bb/2)-
	(B\leftrightarrow\Bb)}{(\cos(sB)-\cos(s\Bb))^2}~
\ee
and similarly for $\tilde{F}(t)$, to give
\bea\label{unrenorm.2} \non
\G^{(2)}_\unren&=&\frac{e^2}{(4\pi)^4}\intz \, W^2\Wb^2\!\!
\int\limits_0^{\infty}\!\!
\rmd s\rmd t\left(F(s,t)-\frac{\tilde{F}(s)}{t}-\frac{\tilde{F}(t)}{s} \right)
\rme^{-\rmi (m^2-{\rm i}\e)(s+t)}\\
&&+2E_1(\rmi m^2s_0)\,
\frac{e^2}{(4\pi)^4}\intz \, W^2\Wb^2\int\limits_0^{\infty}\!\!
\rmd s~ \tilde{F}(s)~\rme^{-\rmi (m^2-{\rm i}\e)s}~.
\eea
Then, motivated by the form of $\G^{(2)}_{\rm ct}$ and by previous 
renormalisations of two-loop Euler-Heisenberg effective actions
we note that
\be \frac{1}{(4\pi)^2}\intz \, W^2\Wb^2\int\limits_0^{\infty}\!\!
	\rmd s~ \tilde{F}(s)~\rme^{-\rmi (m^2-{\rm i}\e) s}
	=m^2\frac{\pd}{\pd m^2}\G^{(1)}~.
\ee

We can now combine $\G^{(2)}_\unren$
with $\G^{(2)}_{\rm ct}$ and choose $Z_Q$ so that the two-loop effective 
action is finite (of course, there is still  freedom in choosing a finite part).
The renormalisation condition then fixes $Z_e$, yielding
\be 
\label{ZQ1}
Z_Q^{(1)}=-\frac{\a}{2\p}E_1(\rmi m^2s_0)~, \qquad
	Z_e^{(2)}=\frac\a\p Z_Q^{(1)}\rme^{-\rmi m^2 s_0}~.
\ee
Thus we see that mass renormalisation at one-loop affects the two-loop 
charge renormalisation.  We note that when calculating with bare parameters 
and multiplicative renormalisation the mass renormalisation contributes to the 
charge renormalisation through the simple relation 
\be\label{massrenorm} \ln(m_0^2)=\ln(m^2)-\ln(Z_m)~. \ee

Since we are using an `on-shell' renormalisation \cite{DR,Coq} 
the appropriate renormalisation equation is the Callan-Symanzik equation 
\cite{CS}.  The renormalisation group functions are defined by
\be \b_{\rm CS}=m\frac{\rmd\a}{\rmd m}=\a\frac{\rmd\ln Z_e}{\rmd\ln m}~,\qquad
	\g_m=\frac{\rmd\ln Z_m^{-\half}}{\rmd\ln m}~.
\ee
In QED it can be shown \cite{DR,Coq} that the $\b$-function for dimensional 
regularisation with minimal subtraction coincides with the above 
$\b$-function to $\ord(\a^3)$.  The proofs given also hold for SQED.
Inspired by (\ref{massrenorm}) we note that to first order 
$Z_Q -1\approx -\half\ln Z_m$,
so to first order in $Z_m$ we can write the charge renormalisation constant as
\be\label{chargerenorm} 
Z_e \approx 1+\frac\a{2\p}\left(\ln(\rmi\rme^\g m^2 s_0)-\ln Z_m\right)~. 
\ee 
It is then a simple calculation to get 
\be \b_{\rm CS}=\frac{\a^2}\p\left(1+\g_m\right)+\ord(\a^3)~,\qquad 
\g_m=\frac\a\p+\ord(\a^2)~.
\ee
These results coincide with the known $\b$ and $\g$ functions, e.g. \cite{SVZ}.
Given that only the one-loop effective action contributes directly to the $F^2$ 
term \cite{SVZ,WSquared} it must be that all higher contributions to the 
charge renormalisation are due to the mass renormalisation.  
Therefore, following the arguments of \cite{SVZ}, we 
expect that (\ref{chargerenorm}) and thus the $\b$-function are exact results.

${}$Finally we can write the renormalised effective action to two loops,
\bea\non
\G[W,\bar{W}]&=&\G^{(0)}+\G^{(1)}+\G^{(2)}=\frac1{e^2}\intc \,W^2 \\
 &&+ \frac{1}{(4\pi)^2}
\intz \,W^2\bar{W}^2\int\limits_0^\infty\!\rmd s
\frac{\Bb^2\left(\cos(sB)-1\right)-B^2\left(\cos(s\Bb)-1\right)}{s\;B^2\Bb^2
\left(\cos(sB)-\cos(s\Bb)\right)}\,\rme ^{-\rmi (m^2 -{\rm i}\e) s} \non \\ 
 &&+\frac{e^2}{(4\p)^4}\intz \, W^2\Wb^2\!\!\int\limits_0^{\infty}\!\!
\rmd s\rmd t\left(F(s,t)-\frac{\tilde{F}(s)}{t}-\frac{\tilde{F}(t)}{s} \right)
\rme^{-\rmi (m^2-{\rm i}\e)(s+t)}~,\qquad
\label{fullEA}
\eea
where $B$ is now understood as
\be
B^2 = \frac{1}{4} D^2 W^2~, 
\ee 
and the vector multiplet is not subject to any constraints.
This is the final form of our main result, and using it allows one to compute, 
by standard means, 
quantities of interest, such as the vacuum non-persistence amplitude 
\cite{Schwinger,Ritus}.

%%%%%%%%%%%%%%%%%%%%%%%%%%%%%%%%%%%%%%%%%%%%%%%%%%%%%%%%%%%%%%%%%%%%%%%%%%%%%%%%
\sect{Self-Dual Background}
In this section we will examine the self-dual limit of the Euler-Heisenberg 
 effective action calculated above.
Six years ago in \cite{Dunne:2001pp,Dunne:2002qf},
it was noted that for both the one and two-loop
Euler-Heisenberg effective actions in scalar and spinor QED the proper-time
integrals could be fully integrated when the background field is self-dual. 
The results of the proper-time integrals can be written  completely in terms 
of the function
\be 
\x(x)=\half\int\limits_0^\infty\!\rmd s\left(\frac1{s^2}-
	\frac1{\sinh^2s}\right)\rme^{-2xs} 
	=-x\left(\psi(x)-\ln x+\frac1{2x}\right)~, 
\ee
where $\j (z) = \G' (z) / \G(z)$ is the digamma function \cite{AandS}.

As discussed in \cite{Kuzenko:2004yd} and references therein, 
the effective action for a supersymmetric theory becomes
trivial in the case of a self-dual background.  Yet we can still impose 
a relaxed form of self-duality which 
allows us to retain a holomorphic-like sector of the effective action.  
If we write the full Euler-Heisenberg 
effective action as in eq. (\ref{structure})
and impose the relaxed self-duality conditions
\be\label{selfdualconds} W_\a \neq 0~, \quad D_\a W_\b = 0~,\quad 
\Db_{\ad}\Wb_{\bd} =	\Db_{(\ad}\Wb_{\bd)} \neq 0 
\ee
then we can track the following sector
\be
 \non \intz W^2\Wb^2 ~ \O(0,
 {\bar D}^2  {\bar W}^2) ~.
  \ee
It should be noted that although the conditions (\ref{selfdualconds}) 
are inconsistent with the structure of 
a single, real vector multiplet, their use is perfectly justified as long 
as we realise we are only calculating the above term.
At the end of the calculation we can remove the self-duality condition 
and have a well defined sector of the effective action.
Since we already have the full two-loop Euler-Heisenberg effective action, 
we can simply take its limit as
$D^2W^2\to 0$ to obtain the above sector.

Taking the self-dual limit of the one-loop effective action we get
\be \non \G_{\rm SD}^{(1)}=\frac{x^2}{(4\p)^2}\intz \frac{W^2\Wb^2}{m^4}
	\int\limits_0^\infty\!\rmd s \,
	s\left(\frac1{s^2}-\frac1{\sinh^2s}\right)\rme^{-2sx}~,
\ee
where we have rescaled $s$ to be dimensionless and written the field strength 
in terms of $x=m^2/\Bb$, a natural dimensionless variable.  
As is shown in \cite{Kuzenko:2004yd},
this can then be expressed in terms of the first derivative of $\x$,
\be 
\G_{\rm SD}^{(1)}=-\frac{1}{(4\p)^2}\intz \frac{W^2\Wb^2}{m^4}~x^2\x'(x)~.
\ee

We will split the two-loop effective action into parts, writing
\be 
\G_{\rm SD}^{(2)}=\frac{e^2}{(4\p)^4}\intz\frac{W^2\Wb^2}{m^4}
\left(I_{I\!I}+I_f+I_\scF \right)~. 
\ee
The first term, $I_{I\!I}$, is the contribution from  Figure \ref{2loopgraphII},
and is calculated, as in \cite{Kuzenko:2004yd},
to be 
\be I_{I\!I} = \third\left(1+x^2\x'''(x)\right)~. \ee

The second term is a bit more difficult, being generated by
\bea 
I_f &=& 2m^4\int\limits_0^{\infty}\!\rmd s\rmd t\left(\frac{P_+P_-}{a_+-a_-}
\frac{f(s)}{s^2t^2}
\ln\left(\frac{a_+}{a_-}\right)-\frac1{t}\tilde F(s)\right)
\rme^{-\rmi m^2(s+t)}\Bigg|_{B\to0} \\\non
&=&4x^2\int\limits_0^{\infty}\!\rmd s\rmd t\frac{s^2}{\sinh^2s}
\left(\coth s-\frac1s\right)
\left(\coth(s+t)-\coth t+\frac1t\right)\rme^{-2x(s+t)}~,
\eea
where we have used the identity
\be \frac{\sinh{t}}{\sinh{s} \sinh(s+t)} = \coth s-\coth(s+t)~. \ee
Part of the above integral for $I_f$ factorises and is easily computed, to yield
\be I_f=-2x^2\int\limits_0^{\infty}\!\rmd s\rmd t\coth(s+t)\frac{\rmd}{\rmd s}
	\left(\frac{s^2}{\sinh^2s}\right)
	\rme^{-2x(s+t)}- 2x^2\x(x)\x''(x)~. \non
\ee
The entangled term can then be simplified by 
repeated integration by parts to give
\be I_f=2x\x(x)- 2x^2\left(\x'(x)+\x(x)\x''(x)\right)-x^3\x''(x)~. \ee

The final term is generated by
\bea
I_\scF&=&m^4\int\limits_0^{\infty}\!
\frac{\rmd s\rmd t}{s^2t^2}\frac{P_+P_-}{a_+-a_-}
\left(\frac{\scF_-}{a_-}-\frac{\scF_+}{a_+}+\frac{\scF_+-\scF_-}{a_+-a_-}
\ln\left(\frac{a_+}{a_-}\right)\right)
\rme^{-\rmi m^2(s+t)}\Bigg|_{B\to0} \\\non
&=&\frac16\left(1+x^2\x'''(x)\right)+2x^2\int\limits_0^{\infty}\!
\frac{\rmd s\rmd t}{\sinh^2(s+t)}
\left(\frac{s^2}{\sinh^2s}-\frac{st\cosh(s+t)}{\sinh s \sinh t}\right)
\rme^{-2x(s+t)}~,
\eea
where we have separated off a term proportional to $I_{I\!I}$.  

We have not been able to compute the remaining double integral analytically, 
however using some techniques standard to experimental mathematics, e.g. 
\cite{exp.math}, 
we have been able to deduce its solution\nolinebreak
\footnote{SJT would like to thank Dr Paul Abbott for suggesting the following 
method and for supplying the initial routine for identifying 
transcendental numbers.}.   
The trick consists of two parts.  Firstly we use high precision numerical 
integration to evaluate the integral for small integer values of $x$ 
corresponding to large field strengths, $\Bb\sim\nolinebreak m^2$.
We then use the hypothesis that the self-dual effective action can always 
be reduced to a combination of derivatives of $\x(x)$ functions with 
polynomial coefficients.  
For any particular value of $x$ these derivatives form an independent 
set of transcendental numbers.
We can then use lattice reduction\nolinebreak
\footnote{Mathematica implements the Lenstra-Lenstra-Lovasz \cite{LLL} 
	algorithm of lattice reduction.}
to find the simplest coefficients that match our numerical integral.
Doing this for a few values of $x$ was enough 
to deduce the functional form of $I_\scF$, the result being
\be 
I_\scF= \frac16\left(1+x^2\x'''(x)\right)+\frac14+x^2-x^3\x''(x)-
\left(\x(x)-x\x'(x)+x\right)^2~.
\ee
This result can be checked by comparing its asymptotic expansion with the 
series expansion of the double integral for weak, self-dual fields using
\be
\int\limits_0^\infty\!\rmd s\rmd t \,\frac{s^nt^m}{(s+t)^l} \,\rme^{-2x(s+t)}
=\frac{n!m!(n+m-l+1)!}{(n+m+1)!(2x)^{n+m+2-l}}~, \quad l\leq n+m+1~.
\ee  
This check is trivial when using computer algebra and has been done to the 
$100^{\rm th}$ order in field strength.

We now can write the full two-loop self-dual effective action as
\bea \G_{\rm SD}^{(2)}&=&\frac{e^2}{(4\p)^4}\intz\frac{W^2\Wb^2}{m^4}
\left(\half\left(1+x^2\x^{'''}(x)\right)
-2x^3\x^{''}(x)+ \right.
\non\\ &&\left.+\quart-2x^2\x(x)\x^{''}(x)
-\left(\x(x)-x\x^{'}(x)\right)^2 \right)~, 
\eea
where we note the first term is just the $\cN=2$ self-dual effective action.  
${}$From here it is easy to read off expansions for both the 
weak ($x\gg 1$) and strong ($x\ll 1$) field limits using \cite{AandS}
\bea
	\x(x)&=&\sum_{n=1}^\infty \frac{B_{2n}}{2nx^{2n-1}}~,\qquad x\gg 1 \\
	\x(x)&=&\half+x(\g+\ln x) - \sum_{n=2}^\infty\z(n)(-x)^{n}~,\quad x\ll 1~,
\eea
where $B_n$ are the Bernoulli numbers and $\g$ is the Euler-Mascheroni constant.
This allows for a much simpler deduction of 
the strong field asymptotics and the particle creation rate
than from the effective action (\ref{fullEA}).

\noindent
{\bf Acknowledgements:}\\
We are grateful to Joseph Buchbinder for reading the manuscript and
useful comments.  
One of us (S.J.T.) is grateful to 
Ian McArthur and Paul Abbott for useful discussions.
The work of S.M.K.\  is supported  in part
by the Australian Research Council. 
The work of S.J.T.\  is supported by an Australian Postgraduate Award.

%%%%%%%%%%%%%%%%%%%%%%%%%%%%%%%%%%%%%%%%%%%%%%%%%%%%%%%%%%%%%%%%%%%%%%%%%%%%%%%%
\begin{appendix}
\sect{Exact superpropagators}
\label{app-superprop}
In this appendix we review, following \cite{KM2}, the structure of the exact 
superpropagators in a  constant $\cN=1$ abelian vector multiplet background.
We start with the gauge covariant derivative algebra defined in 
(\ref{N=1cov-der-al}) 
where the field strengths satisfy the Bianchi identities
\begin{equation}
	\bar\cD_{\ad} \cW_\a = \cD_\a \cWb_{\ad} =0~, 
	\qquad \cD^\a \cW_\a = {\bar \cD}_{\ad} {\bar \cW}^{\ad}~.
\end{equation}

The three major d'Alembertians that occur in covariant supergraphs \cite{GGRS} 
are the vector, chiral and antichiral d'Alembertians, defined by:
\bea \Box_\rmv &=& \cD^a\cD_a-\cW^\a\cD_\a+\cWb_\ad\cDb^\ad~, \non\\
	 \Box_+ &=& \cD^a\cD_a-\cW^\a\cD_\a-\half (\cD^\a\cW_\a)~,\quad
	\Box_+\Phi=\frac{1}{16}\cDb^2\cD^2\Phi~, \quad \cDb_\ad\Phi=0 ~,\\\non
	 \Box_- &=& \cD^a\cD_a+\cWb_\ad\cDb^\ad+\half(\cDb_\ad\cWb^\ad)~, \quad
	\Box_-\bar\Phi=\frac{1}{16}\cD^2\cDb^2\bar\Phi~, \quad \cD_\a\bar\Phi=0  ~.
\eea
The operators $\Box_+$ and $\Box_-$ are related to each other as follows:
\be 
\cD^2\Box_+=\Box_-\cD^2~, \hspace{20pt} \cDb^2\Box_-=\Box_+\cDb^2~,
\ee
whilst for an on-shell background we get the additional, important relations
\be 
\cD^\a\cW_\a=0~\implies~\cD^2\Box_+=\cD^2\Box_\rmv=\Box_\rmv\cD^2~, 
\hspace{10pt} \cDb^2\Box_-=\cDb^2\Box_\rmv=\Box_\rmv\cDb^2~.
\ee
In what follows, the background vector multiplet is chosen 
to be covariantly constant and on-shell,
\be 
\label{cc.os} \cD_a\cW_\a
=0~,\hspace{20pt} \cD^\a\cW_\a=0~. 
\ee

Associated with the d'Alembertian $\Box_\rmv$ is the 
propagator $G(z,z')$ satisfying the equation
\be 
(\Box_\rmv-m^2)G(z,z')=-\delta^8(z-z')~.
\label{greensfn0}
 \ee 
It has the proper-time representation
\be
\label{greensfn} 
G(z,z')=\rmi\int\limits_0^\infty\!\rmd s\, K(z,z'|s)
\rme^{-\rmi(m^2-\rmi\epsilon)s}~,\qquad \epsilon\to+0~. 
\ee
With the corresponding heat kernel \cite{KM2}
\be
\label{heatk} K(z,z'|s)=-\frac{\rmi}{(4\pi s)^2}\sqrt{\det\left(\frac{2s\cF}
{{\rm e}^{2s\cF}-1}\right)}~U\!(s)~
	\zeta^2\bar\z^2\rme^{\frac\rmi{4}\rho\,\cF\!\coth(s\cF)\,\rho}I(z,z')~,
\ee
where the determinant is computed \wrt the Lorentz indices,
\be 
U\!(s)=\exp\left(-\rmi s(\cW^\a\cD_\a-\cWb_\ad\cDb^\ad)\right)~,
\ee
and the supersymmetric two-point functions 
$\zeta^A(z,z')=(\rho^a,\zeta^\a,\bar\z_\ad)$ 
are defined as follows:
\be\label{zeta} 
\rho^a=(x-x')^a-\rmi\theta\sigma^a\bar\theta'
+\rmi\theta'\sigma^a\bar\theta, \quad \zeta^\a
=(\theta-\theta')^\a,\quad \bar\z_{\ad}=(\bar\theta-\bar\theta')_{\ad}~. 
\ee
$I(z,z')$ is the $\cN=1$ parallel displacement propagator described in 
\cite{KM2}.  
The only properties we need for this calculation are
\be
\label{par-prop} I(z,z')I(z',z)=I(z,z)=1 ~.
\ee
Introducing the notation for proper-time dependent variables
$\Psi(s)\equiv U\!(s)\Psi U\!(-s)$
the action of $U\!(s)$ on the objects appearing in the right hand side of 
(\ref{heatk}) 
is\footnote{It should be noted that unlike \cite{KM}, 
the index contractions in spinorial matrix functions will
always be done in the `natural' positions.}
\begin{gather}
\begin{aligned}\non
\cW^\a(s)&=(\cW {\rm e}^{-\rmi s\cN})^\a~,\\
\zeta^\a(s)&=\zeta^\a+(\cW\frac{{\rm e}^{-\rmi s\cN}-1}\cN)^\a~,\quad \\
\end{aligned}
\begin{aligned}
\cWb_{\ad}(s)&=(\cWb {\rm e}^{\rmi s\cNb})_{\ad}~, \\
\bar\z_{\ad}(s)&=\bar\z_{\ad}+(\cWb\frac{{\rm e}^{\rmi s\cNb}-1}{\cNb})_{\ad} 
~,\\ 
\end{aligned}\\
\label{UofsAction}
\begin{align}
\rho_{\a\ad}(s)&=\rho_{\a\ad}-
	2\int_0^s{\rm d}t(\cW_\a(t)\bar\z_\ad(t)+\zeta_\a(t)\cWb_\ad(t))~,\\\non
I(z,z'|s)&=\exp\Big\{\int_0^s{\rm d}t~\Xi(\zeta,\cW,\cWb|t)\Big\}I(z,z')~,
\end{align}
\end{gather}
where
\bea \label{defN} \cN_\a^{~\b}&=&\cD_\a\cW^\b~, 
\qquad \cNb^\ad_{~\bd}=\cDb^\ad\cWb_\bd~,
\eea
and
\bea\label{Xi} \Xi(\zeta,\cW,\cWb)=& &\frac1{12}\rho^{\ad\a}(\cW^\b\bar\z^\bd
	-\zeta^\b\cWb^{\bd})(\eps_{\b\a}\bar\cD_{\bd}\cWb_{\ad}-
	\eps_{\bd\ad}\cD_\b\cW_\a)-\frac{2\rmi}{3}\zeta\cW\bar\z\cWb-\nonumber\\
	&&-\frac\rmi{3}\zeta^2(\cWb^2-\quart\bar\z\bar\cD\cWb^2)
	-\frac\rmi3\bar\z^2(\cW^2-\quart\zeta\cD\cW^2)~.
\eea

Associated with the chiral d'Alembertian $\Box_+$ is the 
propagator $G_+(z,z')$ satisfying the equation
\be\label{gpluseom} (\Box_+ -m^2)G_+(z,z')=-\delta_+(z-z')~,
\hspace{20pt}\delta_+(z-z')=-\quart\cDb^2\delta^8(z-z')~. \ee 
It is covariantly chiral in both arguments,
\be \cDb_\ad G_+(z,z')=\cDb'_\ad G_+(z,z')=0 \ee
and for on-shell backgrounds, $\cD^\a\cW_\a=0$, it is related to $G(z,z')$ via
\be G_+(z,z')=-\quart\cDb^2G(z,z')=-\quart\cDb'^2G(z,z')~. \ee
The corresponding heat kernel is
\be
\label{chheatk} 
K_+(z,z'|s)=-\frac{\rmi}{(4\pi s)^2}
\sqrt{\det\left(\frac{2s\cF}{{\rm e}^{2s\cF}-1}\right)}
~U\!(s)~\zeta^2\rme^{\frac\rmi{4}\rho\,\cF\!\coth(s\cF)\,\rho
-\frac\rmi{2}\rho^a\cW\sigma_a\bar\z}I(z,z')~.
\ee
It is equivalent to the kernels originally computed in \cite{SY,BK85}.
An antichiral Green's function can be similarly constructed.

The other Green's function that occurs in our Feynman rules is the 
antichiral-chiral propagator $G_{-+}$. 
It plays a central role in our calculations, 
and is described in Section \ref{sect2loop}.
%
%%%%%%%%%%%%%%%%%%%%%%%%%%%%%%%%%%%%%%%%%%%%%%%%%%%%%%%%%%%%%%%%%%%%%%%%%%%%%%%%
%
\sect{Two-point function and K\"ahler potential}\label{app.kahler}
The Euler-Heisenberg action constitutes a slowly varying part of the full 
effective action for $\cN=1$ SQED, 
which depends on the vector multiplet only.
Another important sector  of the effective action is  the chiral matter action 
$\G[Q ,Q^\dagger ]$
which is singled out by switching off the gauge field. 
For completeness, we discuss here  the one-loop quantum correction to 
$\G[Q ,Q^\dagger ]$
within the background field method.  
More specifically, we concentrate on evaluating 
the two-point function and the K\"ahler potential. 
Either of these results can then be used to find 
the one-loop renormalisation of the chiral fields.
 
Starting from the classical action (\ref{n=1sqed-action1}) 
we perform the background-quantum splitting
\be V\to e\,v~,\qquad Q_\pm \to Q_\pm + q_\pm~.
\ee
Then, introducing the matrix notation
\be 
\hat{v}=\s_3v~, \qquad \hat{m}=\s_1m~, \qquad q^{\rm T}=(q_+,q_-)~,
\ee
and the gauge invariant quantities
\be
\label{MvandKappa} M_v^2=e^2 Q^\dag Q~, \quad
\k=e^2 Q^{\rm T} \s_1 Q~, \quad
\bar\k=e^2 Q^\dag \s_1 \bar{Q}~,
\ee
the resulting quadratic quantum action takes the form
\be\non
	S_{\rm quad}=-\half\intz~v\left(\Box-M_v^2\right)v 
	+ \intz\left(q^\dag q+eq^\dag\hat{v}Q+eQ^\dag\hat{v}q\right)
	+\half\left(\intc~q^{\rm T}\hat{m}q+\cc\right)~,
\ee
where we have chosen the Fermi-Feynman gauge. 

The mixing terms in $ S_{\rm quad}$ can be 
eliminated from the path integral by implementing the shift
\be 
q(z) ~\to ~q(z)-e\intz'\left(G_{++}(z,z')\hat{v}(z')
\bar{Q(z')}+G_{+-}(z,z')\hat{v}(z')Q(z')\right)~,
\ee
where 
\bea \label{matrixchprop}
{\bf G}&=&\begin{pmatrix} G_{++} & G_{+-} \\ G_{-+} & G_{--}\end{pmatrix} = 
 \begin{pmatrix} \hat{m}\frac{\Db^2}{4} & \frac{\Db^2D^2}{16}\ds1 
 	\\ \frac{D^2\Db^2}{16}\ds1 & \hat{m}\frac{D^2}{4}\end{pmatrix}G ~,\\
\label{basechprop}
 G(z,z')&=&-\frac{1}{\Box-m^2}\d^8(z,z') \non
 \\ &=&
 	\frac{1}{(4\p)^2}\int\limits_0^\infty\frac{\rmd s}{s}\,
\rme^{\frac{\rmi}{4s}\r^2-\rmi s (m^2-{\rm i} \e)}\d^4(\z)
=G_{\rm bos}(z,z') \,\d^4(\z)~,
\eea
are the chiral fields propagators.  
The corresponding Jacobian is obviously equal to
unity.
Then, the  quadratic action turns into 
\be 
S_{\rm 0}=-\half\intz~v\left(\Box-M_v^2-\D\right)v 
+ \intz\,q^\dag q +\half\left(\intc~q^{\rm T}\hat{m}q+\cc\right)~,
\ee
where the operator $\D$, coming from the above field redefinition, is
\bea
\non
 	\D(z,z')&=&e^2\left(Q^\dag(z)\frac{\Db^2D^2}{16}G(z,z')Q(z')
	+Q^{\rm T}(z)\frac{D^2\Db^2}{16}G(z,z')\bar{Q}(z') \right.\\
	&& -\left. Q^\dag(z)\hat{m}\frac{\Db^2}{4}G(z,z')\bar{Q}(z')
	-Q^{\rm T}(z)\hat{m}\frac{D^2}{4}G(z,z'){Q}(z') \right)~.
\eea
Since the components of $\bf G$ in (\ref{matrixchprop}) are 
background independent, the one-loop effective action is 
calculated purely from the gauge field's Hessian,
\be \G_\unren^{(1)}=\frac{\rmi}2\Tr\ln\left(\Box-M_v^2-\D\right)~. \ee

In our first approximation we want to discard all terms that are more than 
quadratic in the chiral background.
This can be achieved by expanding the logarithm to first order,
\be \G_\unren^{(1)}\approx\frac{\rmi}2\Tr\left(\ln(\Box)
-\frac1\Box M_v^2-\frac1\Box\D\right)~. 
\ee
Due to a lack of spinor derivatives to annihilate the Grassmann delta function, 
the first two terms above evaluate to zero.  
Similarly the last two
terms in $\D$ also do not contribute.
This leaves
\be \G_\unren^{(1)}\approx \frac\rmi2e^2\intz\rmd^8z''
G_0(z,z'')\left(Q^\dag(z'')G_{\rm bos}(z'',z')Q(z')
+Q^{\rm T}(z'')G_{\rm bos}(z'',z')\bar{Q}(z') \right)\Big|_{z=z'}\non
\ee
where the Green's function $G_0$ is defined in (\ref{G0}),
and $G_{\rm bos}$ is the bosonic part of (\ref{basechprop}).
Using the proper-time representation for the Green's functions we get
\be\label{2point} 
\G_\unren^{(1)}\approx -\rmi e^2\intz\rmd^4\rho \,Q^\dag(z)Q(z')
\int\limits_0^\infty\frac{\rmd s\, \rmd t}{s^2t^2}\rme^{\rmi\frac{s+t}{4st}\rho^2
-\rmi s (m^2-{\rm i}\e) }\Big|_{\q=\q'}~.
\ee
We note that the above expression 
involves a single Grassmann integral, 
although it is non-local in space-time, in accordance
with the $\cN=1$ non-renormalisation theorem.  It is not difficult to check  
that this is equivalent to the standard
momentum space representation for the two-point function, see, e.g.,
\cite{GGRS,West}.

To compute 
the K\"ahler potential, 
it suffices to choose $Q$ and $Q^\dagger$ to be constant, 
and then $\D$ reduces to
\be 
\D(x,y)=-\frac{1}{\Box-m^2}\left(\frac1{16}M_v^2\{D^2,\Db^2\}-m\k\frac{D^2}4
-m\bar\k\frac{\Db^2}4\right)\d^8(z,z')~, 
\ee
where $\k$ and $\bar\k$ are defined in (\ref{MvandKappa}) above. 
The effective action is then
\be \G_\unren^{(1)}=\frac\rmi2\Tr\ln\Bigg(1+
\frac{1}{16}\,
\frac{
M_v^2 \{D^2,\Db^2\}-4m \k D^2-4 m\bar\k \Db^2  }
	{(\Box-m^2)(\Box-M_v^2)} \Bigg)~.
\ee
The logarithm can then be factorised using
\begin{subequations}\non
\begin{gather}
1+X\Db^2D^2+YD^2\Db^2+ZD^2+\bar{Z}\Db^2=(1+ND^2)(1+U\Db^2D^2
+VD^2\Db^2)(1+\bar{N}\Db^2)\\
N=(1+16\Box Y)^{-1}Z~,\quad V=Y~,\quad
U=X-\bar{Z}(1+16\Box Y)^{-1}Z~, 
\end{gather}
\end{subequations}
for constant, matrix coefficients.
Evaluating the trace in the standard way, by going to momentum space, 
final result for the K\"ahler potential is
\be\label{kahler} 
K^{(1)}_\unren=-\frac\rmi2\int\frac{\rmd^4k}{(2\p)^4}\frac1{k^2}
\ln\left(\frac{(k^2(k^2+m^2)+m^2M_v^2)^2
+k^2m^2\k\bar\k}{k^4(k^2+M_v^2)^2}\right)~. 
\ee
This 
can be compared with the 
calculation given in \cite{Grisaru:1996ve}.
Although we can factorise the above quartic in $k^2$ 
and thus perform the momentum integration, 
it is not very enlightening.
On the other hand, in the massless limit the result is greatly
simplified.  Upon renormalisation at a non-zero field strength
$Q_0$, we get the familiar result (see, e.g., \cite{PW})
\be K^{(1)}\Big|_{m=0}=\frac{e^2}{(4\p)^2}Q^\dag Q
\Big( \ln\frac{Q^\dag Q}{Q_0^\dag Q_0}-2 \Big)~.
\ee
Its functional form is similar to the one-loop K\"ahler potential 
for the Wess-Zumino model first computed in \cite{BKY}.

To renormalise the matter sector 
we only need 
the quadratic part of the K\"ahler potential.
This can be obtained by either 
setting $Q(z') \approx Q(z)$
in (\ref{2point}) or by expanding the logarithm in (\ref{kahler}).
Choosing the latter course gives
\be\non
 K^{(1)}_\unren = \rmi M_v^2\int\frac{\rmd^4k}{(2\p)^4}
 \,\frac1{k^2}\frac{1}{k^2+m^2} + O(|Q|^4)
	=-\frac{M_v^2}{(4\p)^2}\int\limits_{s_0}^\infty\frac{\rmd s}{s}
	\rme^{-\rmi s (m^2-{\rm i}\e)} +O(|Q|^4)~,
\ee
where we have introduced a proper-time cut-off.  
Enforcing the physical renormalisation
condition
\be 
\frac{\pd^2 K}{\pd Q^\dag\pd Q}\Big|_{Q=0}=1, 
\ee
yields the matter renormalisation constant
\be 
Z_Q=1-\frac{\a}{2\p}E_1(\rmi s_0 m^2)+\ord(\a^2) =1-Z_Q^{(1)} +\ord(\a^2) . 
\ee
Clearly, the expression for $Z_Q^{(1)} $ just obtained  coincides with that 
derived in section 5, eq.  (\ref{ZQ1}), on the basis of the two-loop  
renormalisation of the Euler-Heisenberg action.
%
%%%%%%%%%%%%%%%%%%%%%%%%%%%%%%%%%%%%%%%%%%%%%%%%%%%%%%%%%%%%%%%%%%%%%%%%%%%%%%%%
\end{appendix}
%
%%%%%%%%%%%%%%%%%%%%%%%%%  bibliography   %%%%%%%%%%%%%%%%%%%%%%%%%%%%%%%%%%%%%%
\small{

} %end small

\end{document}